\documentclass[conference]{IEEEtran}
\IEEEoverridecommandlockouts

\usepackage{cite}
\usepackage{amsmath,amssymb,amsfonts}
\usepackage{algorithmic}
\usepackage{graphicx}
\usepackage{textcomp}
\usepackage{xcolor}
\usepackage{booktabs}
\usepackage{subcaption}
\usepackage{hyperref}
\usepackage[most]{tcolorbox}
\usepackage{diagbox}
\usepackage{tikz}
\usepackage{xcolor}

\newcommand{\rev}[1]{#1}

\makeatletter 
\newcommand{\linebreakand}{%
  \end{@IEEEauthorhalign}
  \hfill\mbox{}\par
  \mbox{}\hfill\begin{@IEEEauthorhalign}
}
\makeatother 

\def\BibTeX{{\rm B\kern-.05em{\sc i\kern-.025em b}\kern-.08em
    T\kern-.1667em\lower.7ex\hbox{E}\kern-.125emX}}
\begin{document}

\title{From Release to Adoption: Challenges in Reusing Pre-trained AI Models for Downstream Developers\\
}


\author{\IEEEauthorblockN{Peerachai Banyongrakkul}
\IEEEauthorblockA{
\textit{The University of Melbourne}\\
Melbourne, Australia \\
pbanyongrakk@student.unimelb.edu.au}
\and
\IEEEauthorblockN{Mansooreh Zahedi}
\IEEEauthorblockA{
\textit{The University of Melbourne}\\
Melbourne, Australia \\
mansooreh.zahedi@unimelb.edu.au}
\and
\IEEEauthorblockN{Patanamon	Thongtanunam}
\IEEEauthorblockA{
\textit{The University of Melbourne}\\
Melbourne, Australia \\
patanamon.t@unimelb.edu.au}
\and
\linebreakand
\IEEEauthorblockN{Christoph	Treude}
\IEEEauthorblockA{
\textit{Singapore Management University}\\
Singapore \\
ctreude@smu.edu.sg}
\and
\IEEEauthorblockN{Haoyu	Gao}
\IEEEauthorblockA{
\textit{The University of Melbourne}\\
Melbourne, Australia \\
haoyug1@student.unimelb.edu.au}
}

\maketitle

\begin{abstract}
Pre-trained models (PTMs) have gained widespread popularity and achieved remarkable success across various fields, driven by their groundbreaking performance and easy accessibility through hosting providers. However, the challenges faced by downstream developers in reusing PTMs in software systems are less explored. To bridge this knowledge gap, we qualitatively created and analyzed a dataset of 840 PTM-related issue reports from 31 OSS GitHub projects. We systematically developed a comprehensive taxonomy of PTM-related challenges that developers face in downstream projects. Our study identifies seven key categories of challenges that downstream developers face in reusing PTMs, such as model usage, model performance, and output quality. We also compared our findings with existing taxonomies. Additionally, we conducted a resolution time analysis and, based on statistical tests, found that PTM-related issues take significantly longer to be resolved than issues unrelated to PTMs, with significant variation across challenge categories. We discuss the implications of our findings for practitioners and possibilities for future research.

\end{abstract}


\section{Introduction}
In recent years, AI-based software systems are rapidly spreading in our society, fueled by advancements in Deep Learning (DL) \cite{Silverio2022}. The development of AI-based software systems differs greatly from traditional software, affecting all stages from requirements to delivery. This is because these systems present greater complexity, involving new stakeholders (e.g., AI developers) with unique concerns (e.g., model accuracy, data quality) \cite{Muccini2021, Gao24}. To speed up DL development and reduce costs, pre-trained models (PTMs) are increasingly being used and integrated into systems \cite{Han2021, Davis2023}. PTMs are deep learning models trained on large datasets and widely reused across diverse domains, such as GPT-4 \cite{OpenAI2023}. 

\rev{However, reusing PTMs within software systems introduces new forms of complexity beyond traditional software dependencies \cite{Jiang2024}. PTMs are often treated as opaque components, with minimal documentation \cite{Gao24}, unclear behavior \cite{Burrell2016}, lack of standardized versioning \cite{Ajibode2025}, and significant computational demands \cite{Lai2024}. As a result, developers who are not involved in model training or architecture design often encounter difficulties in using models for their purposes \cite{Pan2022, Taraghi2024, Tan2024}. We refer to these developers as downstream developers—that is, practitioners who reuse PTMs in their own software (i.e., downstream software) without directly contributing to the model's development. These developers differ from model creators or ML researchers; they primarily work in software engineering roles and adapt PTMs to fit within their downstream software environments and constraints.}



\rev{Prior studies have focused on various challenges \cite{Davis2023, Jiang2023, Taraghi2024, Tan2024} and bugs \cite{Chakraborty2021, Pan2022} associated with PTMs, but not from the perspective of downstream developers.} Davis et al. \cite{Davis2023} present a vision paper on PTM reuse challenges without empirical validation, while Jiang et al. \cite{Jiang2023} and Taraghi et al. \cite{Taraghi2024} \rev{focus on the Hugging Face ecosystem, emphasizing limitations within upstream communities.} Similarly, Tan et al. \cite{Tan2024} analyze PTM-related issues from Stack Overflow discussions, \rev{which are often disconnected from real-world development contexts.} Meanwhile, Chakraborty et al. \cite{Chakraborty2021} and Pan et al. \cite{Pan2022} examine bugs within the development of PTMs, rather than challenges in downstream projects. Despite these valuable contributions, little is known about the software development challenges faced by downstream software projects that reuse PTMs from the developer's perspective. \rev{To bridge this gap, our study is the first to empirically investigate the PTM-reuse-specific challenges reported by software practitioners in downstream projects on GitHub, where PTMs are reused in actual development environments.}

We curated a set of downstream software projects on GitHub that employ PTMs. To investigate the challenges associated with PTM reuse, we systematically extracted issue reports from these repositories. \rev{In total, we collected 8,373 issue reports across 31 repositories. Through manual annotation, we identified 840 issue reports} related to PTM reuse and thematically analyzed these issues to understand the challenges. Beside constructing a taxonomy of challenges, we also analyzed and compared the resolution times of PTM and non-PTM issues. In addition, we compared and discussed our results with existing relevant taxonomies and related work. 



Our analysis indicates that downstream developers face challenges in projects reusing PTMs across seven key categories. These include issues related to model usage, performance, software environment, documentation, hardware constraints, resource configuration, and functionality enhancements. Notably, most of these challenges are not mentioned in previous work, highlighting the novelty of our taxonomy. Our taxonomy uncovers critical gaps in several aspects, especially in model functionality enhancement and resource configuration. Furthermore, our findings show that PTM issues take significantly longer to be fixed than non-PTM issues, particularly in performance optimization and functionality enhancements. The replication package is available at Zenodo \cite{anonymous_2025_15020622}.

The key contributions of our research are as follows: 1) a fine-grained taxonomy of challenges when reusing PTMs in downstream software projects, 2) an analysis of resolution times of PTM-related vs. non-PTM-related challenges, 3) a comparative analysis of our results with existing taxonomies and related work, and 4) a replication package with a dataset of 8,373 manually annotated issues from 31 OSS projects reusing PTMs and our qualitative analysis to facilitate future research.

\section{Background and Related Work}
\label{sect:background_related_work}

\subsection{Pre-Trained Model Reuse}
Pre-trained models (PTMs) are deep learning models trained on large datasets and made available for reuse across various applications. PTMs have demonstrated significant success across diverse domains, including software engineering\cite{Karmakar2021}. Popular examples of PTMs include BERT \cite{bert}, GPT-4 \cite{OpenAI2023}, Deepseek \cite{DeepSeek-AI2025}, and Stable Diffusion \cite{sd}. Similar to software package reuse, PTMs are widely adopted in software projects and are distributed via model registries like Hugging Face, TensorFlow Hub, PyTorch Hub, and the ONNX Model Zoo. Hugging Face stands out as the largest and most diverse repository \cite{Jiang2023, Castano2023}. Despite the benefits, reusing PTMs presents its own set of challenges. The black-box nature of these models often leads to software issues, as downstream users may lack a deep understanding of how the model processes data and produces outputs \cite{Pan2022, Taraghi2024}, as well as the infrastructure and resources required to run them \cite{Zhou2023}. 

Therefore, several recent studies have explored these challenges. The work of Davis et al. \cite{Davis2023} envisioned that challenges of PTM reuse would be categorized into three paradigms, including conceptual, adaptation, and deployment reuse, and emphasized standardization and portability concerns. Moreover, Jiang et al. \cite{Jiang2023} and Taraghi et al. \cite{Taraghi2024} examined PTM reuse in the Hugging Face ecosystem, identifying key challenges such as missing attributes, performance discrepancies, training pipelines, and memory constraints. Tan et al. \cite{Tan2024} studied PTM-related discussions on Stack Overflow, revealing challenges in model life cycle management, fine-tuning, and output interpretation. Prior studies also examined bugs related to reusing and the development of PTMs \cite{Chakraborty2021, Pan2022}. Chakraborty et al. \cite{Chakraborty2021} analyzed Stack Overflow posts on BERT, identifying common bugs about fairness issues, parameter misconfigurations, and version incompatibilities. Pan et al. \cite{Pan2022} investigated bugs by mining GitHub repositories for PTM development, finding memory-related and initialization bugs to be prevalent due to PTMs’ large size, along with API misuse and retraining errors. The black-box nature of PTMs further exacerbates the difficulties of debugging. 

While these studies provide valuable insights into PTM reuse, they primarily focus on challenges from the perspectives of platform users, along with forum discussions \cite{Chakraborty2021, Jiang2023, Taraghi2024, Tan2024}, or model repositories \cite{Pan2022}, capturing high-level or conceptual challenges. Little is known about the challenges faced by downstream software projects that integrate PTMs from a developer’s perspective. Our study fills this gap by focusing on GitHub issue reports from downstream software projects reusing PTMs, which provide an empirical understanding of PTM-related challenges faced by software engineers in real-world development environments.

\subsection{Mining GitHub Issues}
Recent empirical studies have analyzed GitHub issues to better understand DL frameworks. Makkouk et al. \cite{Makkouk2022} examined performance bugs in DL frameworks (e.g., TensorFlow, PyTorch), noting their complexity and extended resolution time. Long et al. \cite{Long2022} found that many performance and accuracy issues are unclassified or unrelated to actual bugs, with only half leading to direct fixes. Yang et al. \cite{Yang2022} proposed a taxonomy of DL framework bugs, identifying model-building bugs as a major challenge requiring specialized fixes. Chen et al.\cite{Chen2023} further investigated bugs across four popular DL frameworks by offering a detailed analysis of root causes and symptoms, and providing actionable guidelines to improve bug detection and debugging practices.
In addition, some researchers have focused on issues in downstream AI-based systems \cite{Humbatova2020, Yang2023, Zhang2023, Idowu2024, Lai2024}. Humbatova et al. \cite{Humbatova2020} developed a taxonomy of challenges in conventional DL systems by analyzing GitHub and Stack Overflow, identifying training-related issues, model-specific faults, and API problems. Yang et al. \cite{Yang2023} analyzed open-source AI research projects, highlighting runtime errors and unclear instructions. Zhang et al. \cite{Zhang2023} examined architecture decisions in AI systems, identifying six linguistic patterns. More recently, Idowu et al. \cite{Idowu2024} focused on development stages and evolution in ML-related Python projects. Lastly, Lai et al. \cite{Lai2024} adapted Humbatova et al.'s taxonomy \cite{Humbatova2020} to analyze ML-applied software projects, comparing ML and non-ML issues in terms of fix duration and size. They found that ML issues take longer to be resolved but show no significant difference in fix size.


These studies provide valuable insights into issue management in DL frameworks and downstream AI-based projects, but they primarily focus on challenges encountered when building traditional ML/DL models from scratch using AI frameworks. They overlook the unique challenges of PTM reuse in downstream projects, where models are integrated, adapted, and maintained, rather than developed from the ground up. This gap highlights the need for an empirical study focusing on PTM-related issues within real-world software projects reusing PTMs.


\subsection{Motivating Example}

\begin{figure}
    \centering
    \includegraphics[width=\linewidth]{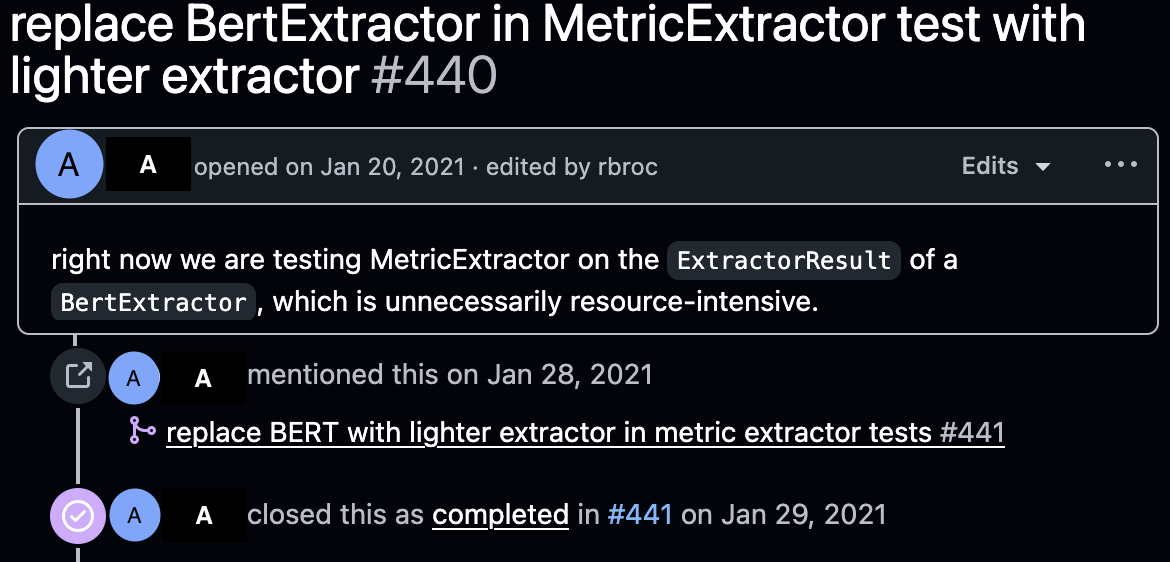}
    \caption{A motivating example illustrating a PTM-related challenge found in downstream software projects.}
    \label{fig:motivation}
\end{figure}

In this study, we focus on the challenges downstream developers face when reusing PTMs in real-world software projects. To illustrate these challenges, we present a case from Pliers\footnote{\url{https://github.com/PsychoinformaticsLab/pliers}\label{Pliers}}, a software project for developing a Python package for automated multimodal extraction of features from different types of media, such as text, image, audio, and video. In this project, the developers integrated BERT into their feature extraction pipeline to analyze textual data. In the course of development, while BERT offers powerful feature extraction, its high computational cost introduce unnecessary computational overhead, slowing down the pipeline. To address this, a developer created \textsl{\href{https://github.com/PsychoinformaticsLab/pliers/issues/440}{Pliers \#440\cite{pliers440}}} to express the need to replace BERT with a lighter model (i.e., VADER),  as shown in Figure \ref{fig:motivation}. This issue reflects a decision-making challenge in PTM integration where developers must balance model capability and resource constraints while developing software. Another example is \textsl{\href{https://github.com/PsychoinformaticsLab/pliers/issues/318}{Pliers \#318\cite{pliers318}}}, where a developer proposed integrating spaCy's PTMs to support their software functionalities. This issue exemplifies the demand for supporting additional models within existing software infrastructures. Integrating new PTMs necessitates careful consideration of compatibility, dependency management, and the potential need for architectural adjustments to accommodate expanded functionalities. These types of challenge do not fit within existing taxonomies, which mainly focus on training-related issues, software environment incompatibilities, model initialization difficulties, and model API usage problems. This demonstrates that PTM reuse in downstream software projects is not only about selecting and applying models, but also involves managing resource constraints (\textsl{\href{https://github.com/PsychoinformaticsLab/pliers/issues/440}{Pliers \#440\cite{pliers440}}}), ensuring compatibility (\textsl{\href{https://github.com/PsychoinformaticsLab/pliers/issues/318}{Pliers \#318\cite{pliers318}}}), and handling versioning issues (\textsl{\href{https://github.com/brycedrennan/imaginAIry/issues/326}{imaginAIry \#326\cite{imaginary326}}}).









\section{Study Design}
\label{sect:study_design}

In this section, we present our research methodology and the study phases, as can be seen in Figure \ref{fig:overview}.

\subsection{Research Questions}

\noindent \textbf{RQ1:} \textit{What PTM-related challenges do downstream developers encounter in software systems?}

Reuse of PTMs comes with its own challenges, spanning both technical limitations and engineering practices \cite{Davis2023}. This research question focuses on identifying and categorizing these challenges to better understand the nature of difficulties downstream developers face in their software systems that reuse PTMs.



\noindent \textbf{RQ2:} \textit{How does the resolution time of PTM-related issues compare to non-PTM-related issues in software systems?}

Examining the resolution time of PTM-related issues is essential to quantify their impact. Building on RQ1, this research question examines whether PTM-related issues take longer to be resolved than non-PTM issues, inspired by \cite{Lai2024}. Investigating these differences helps determine if PTM-specific adaptations of software engineering best practices are needed. This question also explores whether different categories of PTM-related challenges influence resolution time.




\subsection{Data Collection and Pre-processing}
For this study, we utilized the PeaTMOSS dataset \cite{Jiang2024}, which contains 28,575 GitHub projects that reuse 281,638 PTMs from Hugging Face and PyTorch Hub. The data was collected as of July 10, 2023. However, it is possible that these GitHub projects may not be software projects (e.g., PTM repositories). Since our goal is to understand the PTM-related issues in the downstream software projects that reuse PTMs, we conduct the following steps to curate the dataset.

\begin{figure}
    \centering
    \includegraphics[width=\linewidth]{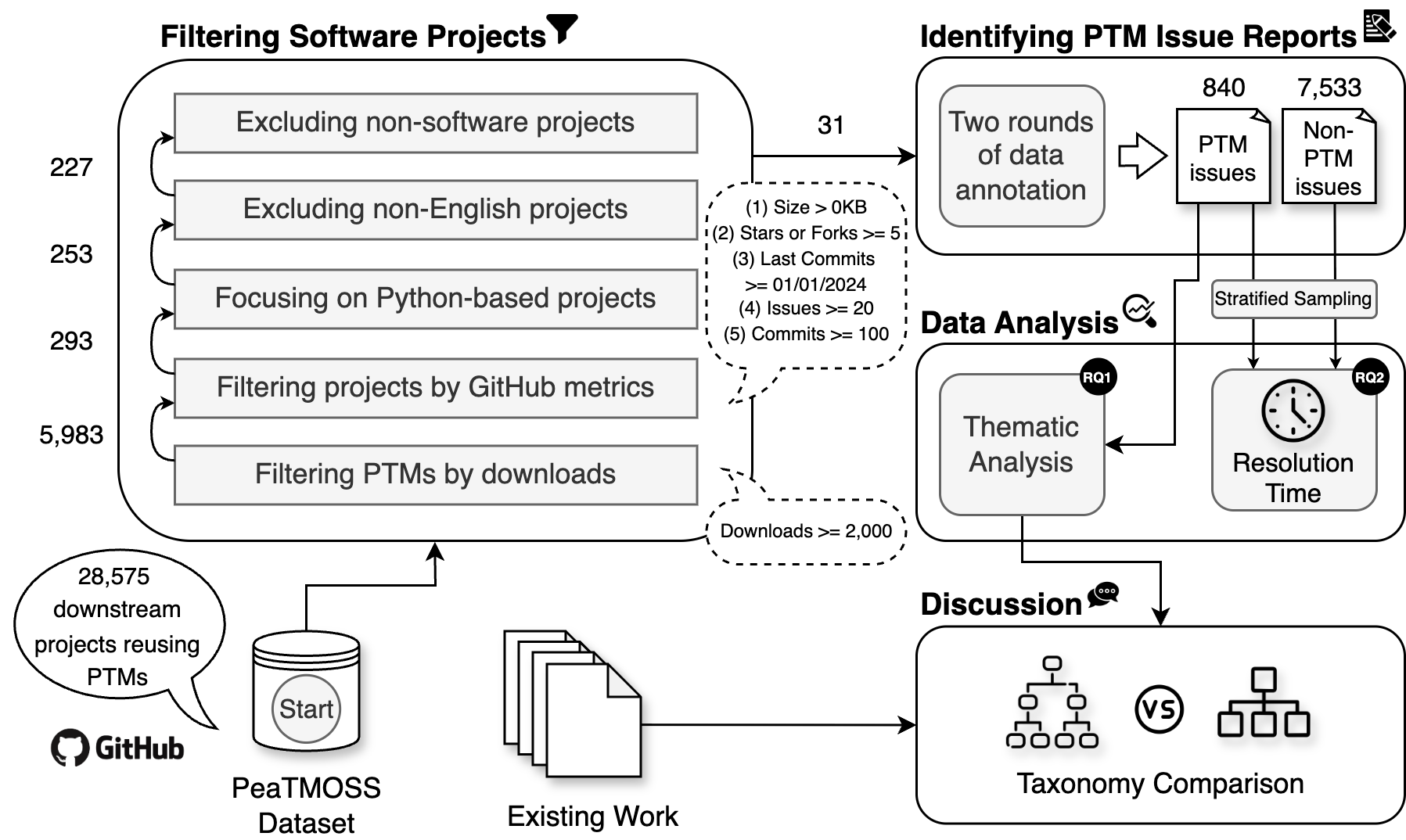}
    \caption{Overview of our study approach.}
    \label{fig:overview}
\end{figure}

\textbf{Filtering Software Projects:}
To ensure the quality and relevance of  data, we filtered the dataset in the following steps: 
Firstly, aiming to investigate the challenges of commonly-used PTMs, we included PTMs with more than 2,000 downloads as suggested by previous studies \cite{Jiang2024, Pepe2024, Castano2023, Jiang2023b}\rev{, reducing the dataset to 5,938 projects and 197,882 issues.} Secondly, we applied GitHub-based metrics, as seen in Figure \ref{fig:overview}, to ensure the selection of active, well-maintained, and relevant projects. These include size, popularity, activity, data availability, content filtering (excluding tutorials, homework, coding challenges, or resource collections via keyword-based filtering from \cite{Toma2024}), issues, and commits. These criteria are adapted from best practices in software repository analysis \cite{Gonzalez2020, Obrien2022, Toma2024, Pan2022, Gao2023, Banyongrakkul2024}. Notably, the thresholds in both steps were chosen based on elbow points, where the curve levels off when plotting each metric against the number of remaining projects. \rev{This resulted in 293 projects and 116,658 issues.} Then, we included only projects written in Python, as it is the most popular programming language for AI-related projects \cite{Gonzalez2020, Biswas2022} and the predominant language in the dataset\rev{, yielding 253 projects and 101,596 issues.} After that, we manually inspected and excluded projects where most content was in a non-English language, such as \textsl{Bark-Voice-Cloning}\footnote{\url{https://github.com/KevinWang676/Bark-Voice-Cloning}}\rev{, resulting in 227 projects and 93,061 issues.}

Finally, we excluded non-software projects through manual validation. Specifically, this involved inspecting repository descriptions, READMEs, and structures to include only projects reusing PTMs as components for their core functionality, targeting non-AI developers or engineers, and not PTM repositories themselves. Firstly, we removed non-technical projects (e.g., research papers, books). Moreover, repositories dedicated to the development of PTMs (e.g., \textsl{ConvLab-3}\footnote{\url{https://github.com/ConvLab/ConvLab-3}}), or repositories of AI tools specifically designed for AI developers, like model benchmarking, training, or deployment (e.g., \textsl{chitra}\footnote{\url{https://github.com/aniketmaurya/chitra}}). To ensure rigor and consensus, manual inspections were performed collaboratively by two authors. Specifically, the first author randomly selected a sample of 50 projects and shared them with another author. Two authors independently annotated them, achieving a Cohen’s kappa of 0.65, indicating substantial agreement \cite{McHugh2012}. Disagreements were resolved collaboratively before applying the criteria to the remaining records by the first author. \rev{The disagreements often arose in borderline cases. One example of a grey-area case is \textsl{FormFyxer}\footnote{\url{https://github.com/SuffolkLITLab/FormFyxer}}, which was initially debated as it is developed by a research group. After discussion, we agreed to include it, as it demonstrates real-world software functionality and clear evidence of software engineering practice.} Ultimately, 37 projects \rev{with 8,671 issues} remained, encompassing a variety of software. For example, AI-powered applications (e.g., image and video generation\footnote{\url{https://github.com/brycedrennan/imaginAIry}}), domain-specific AI-based tools (e.g., medical concept annotation\footnote{\url{https://github.com/CogStack/MedCAT}}), and frameworks and libraries (e.g., a Python package for feature extraction\footref{Pliers}).

\begin{table}[h!tb]
\scriptsize
\centering
\caption{Statistics of the selected downstream repositories}
\label{tab:dataset}
\begin{tabular}{l|rrrr}
\toprule
\multicolumn{1}{l}{\backslashbox{\textbf{Metric}}{\textbf{Statistics}}} & \multicolumn{1}{|c}{\textbf{Min}} & \multicolumn{1}{c}{\textbf{Max}} & \multicolumn{1}{c}{\textbf{Mean}} & \multicolumn{1}{c}{\textbf{SD}} \\
\toprule
\textbf{Stars} & 11 & 26,230 & 3,219.52 & 6,377.21 \\
\textbf{Forks} & 1 & 2,571 & 384.00 & 732.17 \\
\textbf{Commits} & 190 & 6,426.00 & 1,046.57 & 1,236.87 \\
\textbf{Issues} & 28 & 1,871 & 201.00 & 383.25 \\
\textbf{PTM-Related Issues} & 1 & 239 & 27.10 & 48.98 \\
\bottomrule
\end{tabular}
\end{table}

\textbf{Identifying PTM-Related Issue Reports:}
With the filtered software projects, we conducted two rounds of data annotation to identify PTM-related issue reports. The first author randomly selected 100 issues and shared them with another author. Both, with more than four years of ML/DL experience, independently labeled each issue as `PTM' or `Non-PTM' based on the title, description, and discussions. The first round yielded a Cohen’s kappa of 0.60 (moderate agreement) \cite{McHugh2012}. After resolving conflicts, we refined our inclusion/exclusion criteria. In the second round, another 100 reports were annotated using the revised criteria, improving Cohen’s kappa to 0.82 (strong agreement) \cite{McHugh2012}. The remaining disagreements were resolved, further refining the criteria. \rev{Our inclusion criteria explicitly encompass diverse issue types (e.g., bugs, enhancement requests, feature discussions, and general questions), as long as they reflect challenges that developers face related to PTM reuse. These PTM-related issues were identified within three main components of AI-based software: the model itself, the data pipeline, and the training processes, inspired by \cite{Davis2023, Muccini2021}. Additionally, we included issues that discuss AI practices and supporting elements such as software environment, system infrastructure, and documentation.} Exclusion criteria filter out non-PTM issues, including generic bugs, non-model components, research content, non-English issues, and issues with insufficient detail. After reaching consensus, the first author applied the finalized criteria to annotate the remaining issues. Projects without PTM-related issues were removed, resulting in 840 PTM-related issue reports \rev{and 7,533 non-PTM issues} across 31 projects. Table \ref{tab:dataset} presents a descriptive summary of key statistics for the dataset. This dataset \cite{anonymous_2025_15020622} provides a diverse foundation for analyzing PTM reuse in different software development environments.



\subsection{Data Analysis}
In this section we elaborate on the process of analyzing data to address our research questions.

\textbf{Thematic Analysis (RQ1):}
After completion of data preparation, we qualitatively analyzed \rev{840 PTM-related issue reports} using thematic analysis \cite{Cruzes2011, Braun2006} to establish our fine-grained taxonomy of PTM-related challenges. All analyses were performed using NVivo\footnote{\url{https://lumivero.com/products/nvivo}}. \rev{Initially, to become familiar with the data, the first author randomly selected a pilot sample of 50 issue reports for manual inspection with another author. During this phase, we extracted keypoints of challenges that developers encountered, labeled them with preliminary descriptive tags (i.e., codes), and categorized them accordingly. Through this step, we established common understanding to ensure consistency in our data analysis. For instance, a single issue report could contain one or multiple distinct challenges. We also excluded challenges that were too vague, overly broad, or lacked sufficient information for meaningful analysis. These initial codes were then reviewed and discussed by all authors to align the coding scheme and ensure consistency before proceeding with the full dataset.}

\begin{figure*}[!htb]
    \centering
    \includegraphics[width=\linewidth]{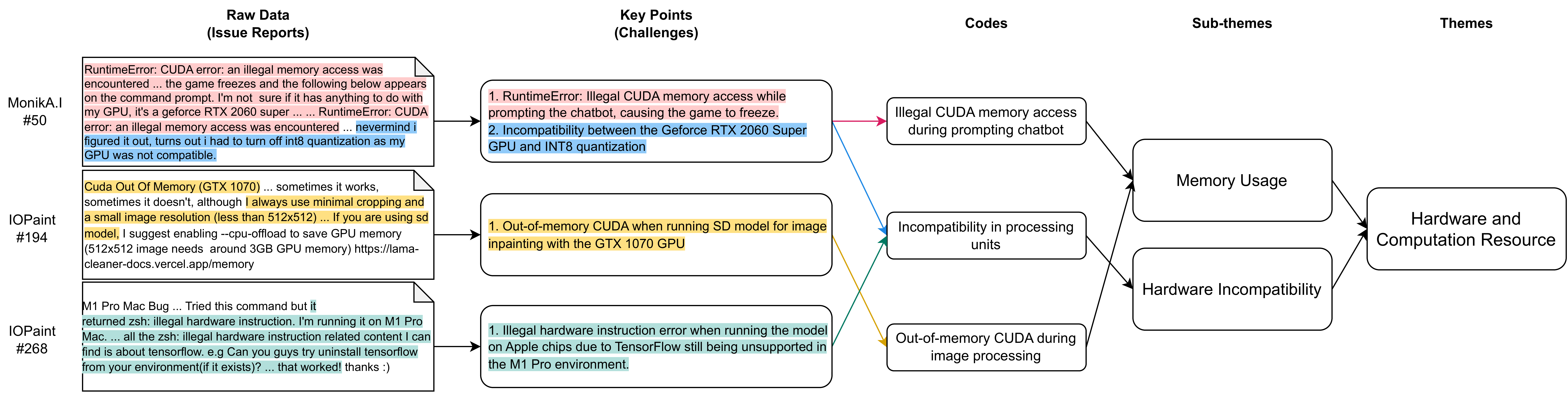}
    \caption{An example of the data analysis process for developing the taxonomy of PTM-related challenges.}
    \label{fig:data_analysis}
\end{figure*}

\rev{Throughout the full analysis, w}e employ a purely bottom-up methodology, iteratively comparing new codes with existing ones to refine and adjust categories based on emerging themes, as exemplified in Figure \ref{fig:data_analysis}. The first author was primarily responsible for this process, which involved the following key steps. Firstly, we systematically reviewed issue reports by analyzing their titles, descriptions, and discussions, with additional reference to related issues and pull requests where applicable. Secondly, we extracted challenges and assigned leaf-level codes to them. Finally, emerging codes were iteratively grouped into higher-level themes \rev{(i.e., sub-themes and themes)} to capture overarching patterns. To maintain quality and consistency, the assignment and categorization of the codes were reviewed and revised every 50 issue reports. All the other authors closely monitored the entire process through weekly meetings. This iterative approach allowed for continuous refinement, where codes were adjusted, merged, split, or removed as needed. In cases of uncertainty or disagreement, discussions were held to reach a final consensus, with conflicts being systematically resolved in regular meetings.

\textbf{Resolution Time Analysis (RQ2):}
\rev{Our objective in this analysis is to determine whether PTM-related issues require more time to resolve than non-PTM issues.} Resolution time is defined as the number of days from creation of an issue to its closure. If an issue is closed on the same day it is opened, its resolution time is zero. \rev{To enable a fair comparison, we first constructed two balanced groups of closed issue reports—PTM-related and non-PTM—by employing stratified sampling \cite{Baltes2022}. For each project, we randomly selected an equal number of closed non-PTM issues to match the number of closed PTM-related issues. If a project had fewer available non-PTM issues, we proportionally reduced the number of PTM issues to maintain balance. Projects where one group had no closed issues were excluded. This process resulted in a dataset of 802 issue reports—401 PTM-related and 401 non-PTM-related—across 27 projects.}


\rev{To further mitigate potential confounding effects, particularly those stemming from differences in issue types, we used a fine-tuned seBERT model \cite{Trautsch2023} to classify issues into three categories: bug, question, and enhancement. These are the predefined issue types supported by the model, and the classification was used solely to provide a high-level overview of issue type distribution and validate the balance between the PTM and non-PTM groups. To strengthen the validity of this classification, we conducted a broader manual validation. Using a statistically representative sample of 260 issues (130 PTM and 130 non-PTM) with 95\% confidence level and 5\% margin of error, we found that 249 cases (95.77\%) matched the manual classification, confirming the reliability of the seBERT-based issue type tagging. A Chi-square independence test \cite{McHugh2013} confirmed that the distribution of issue types did not significantly differ between the groups (p-value = 0.2), supporting the comparability of the two sets.}


\rev{For the statistical comparison of resolution times}, we employed the one-sided Mann-Whitney U Test \cite{Mann1947}, a non-parametric test, with significance level of 0.05 to determine whether PTM-related issues take significantly longer to resolve than non-PTM issues. Additionally, we conducted a Kruskal-Wallis H-test \cite{Kruskal1952}, a non-parametric test, to determine if the medians of PTM-related challenge themes differ, using the same significance level as the previous test. We also computed the average resolution time for PTM-related issues across different taxonomy themes and sub-themes. This allowed us to analyze which challenge categories required the longest resolution time.

\section{Taxonomy of Challenges (RQ1)}
\label{sect:results} 

\begin{figure*}[t]
    \centering
    \includegraphics[width=\linewidth]{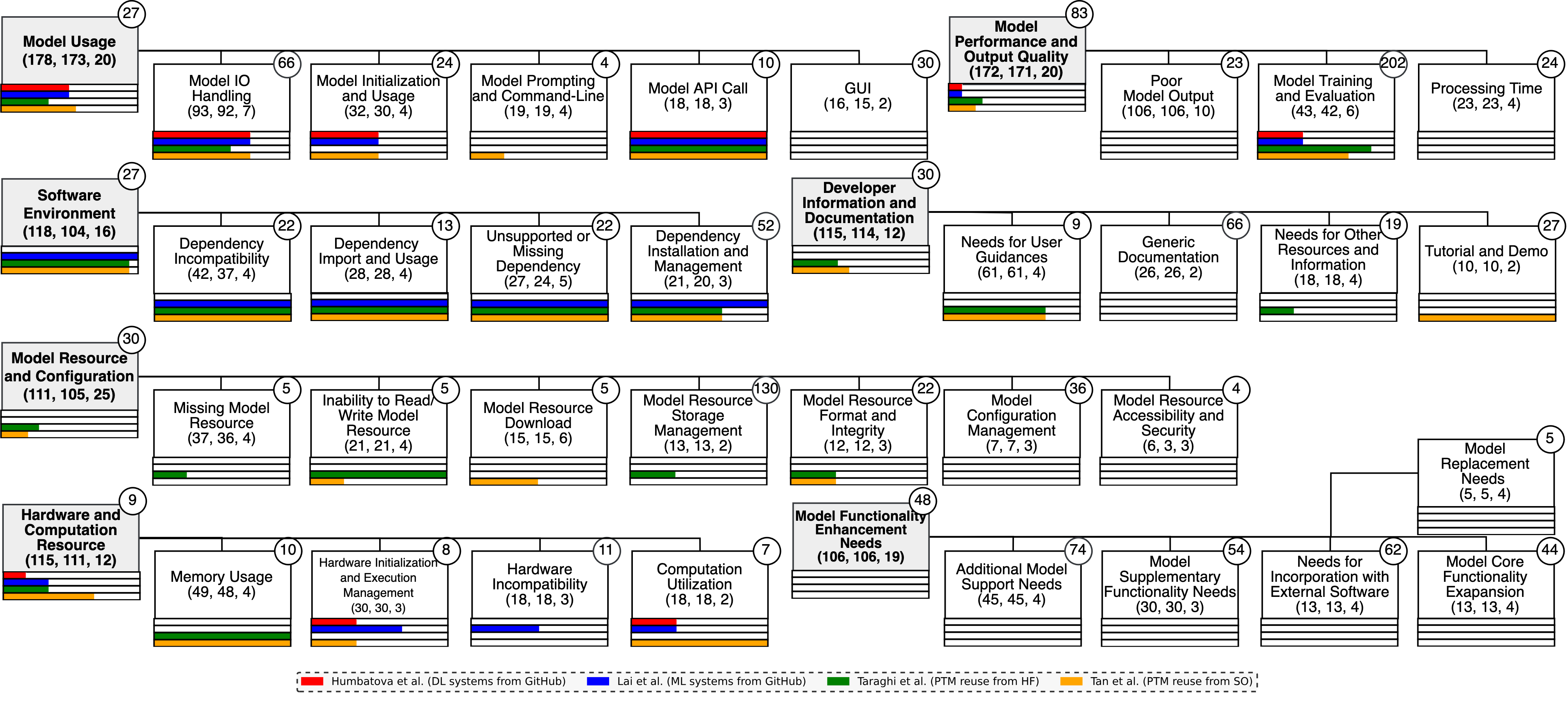}
    \caption{The top two levels of our taxonomy of PTM-related challenges in downstream software projects, where each challenge group is labeled as (\# challenges, \# issue reports, \# codes). The colored bars represent the overlap with existing taxonomies \cite{Humbatova2020, Lai2024, Taraghi2024, Tan2024}, and empty bars indicate new categories. The small circles at the top-right indicate the average resolution time.}
    \label{fig:taxonomy}
\end{figure*}

In this section, we present our final taxonomy of PTM-related challenges. The taxonomy is structured into three hierarchical levels \cite{anonymous_2025_15020622}. Figure \ref{fig:taxonomy} shows the seven top-level themes (gray boxes) including: (1) Model Usage, (2) Model Performance and Output Quality, (3) Software Environment, (4) Developer Information and Documentation, (5) Hardware and Computation Resource, (6) Model Resource and Configuration, and (7) Model Functionality Enhancement Needs. These themes are further divided into 32 sub-themes (white boxes). \rev{In the figure, each challenge group is labeled in the format (number of challenges, number of issue reports, number of codes). For instance, the group Model IO Handling is annotated as (93, 92, 7), indicating that this sub-theme encompasses 93 individual challenge instances drawn from 92 issue reports, and these were abstracted into 7 unique codes.}

\subsection{Model Usage (MU)}
This theme is the most frequent in our taxonomy, accounting for 178 challenges. It covers issues related to interacting with PTMs. We elaborate on the sub-themes of this theme below:

\textbf{Model I/O Handling} (93 challenges): Developers usually face difficulties in handling model input and output, including passing parameters as tensor inputs through a model. One of the most common issues involves mismatches in types or shapes of data or tensors, such as \textsl{\href{https://github.com/threestudio-project/threestudio/issues/260}{Threestudio \#260 \cite{threestudio260}}}. Another frequent challenge is saving model outputs, exemplified in \textsl{\href{https://github.com/neuralchen/SimSwap/issues/133}{SimSwap \#133\cite{simswap133}}}. Model I/O issues also arise from input/output formatting, such as image padding errors and JSON conversion failures. Encoding and tokenization inconsistencies further contribute to these challenges, including embedding mismatches and handling unknown tokens. Developers also struggle with refining model outputs, underscoring the need for systematic post-processing, particularly for string matching and misspelling corrections.



\textbf{Model Initialization and Usage} (32 challenges):
Developers face compatibility issues, missing attributes, and structural mismatches when initializing PTMs. Most issues arise during model training and functionality usage from the user side, such as \textsl{\href{https://github.com/neuralchen/SimSwap/issues/416}{SimSwap \#416\cite{simswap416}}}, indicating a missing model attribute that prevents initialization. Structural mismatches, like hidden layer discrepancies, like \textsl{\href{https://github.com/neuralchen/SimSwap/issues/251}{SimSwap \#251\cite{simswap133}}}, also cause errors.

\textbf{Model Prompting and Command-Line} (19 challenges):
Some software enables user interaction with models via prompts and CLI, but developers often encounter challenges in handling command-line parameters, such as missing model options (\textsl{\href{https://github.com/Sanster/lama-cleaner/issues/90}{IOPaint \#90\cite{iopaint90}}}). Formatting issues also arise, including excessive technical details in prompt templates and handling malformed inputs. Additionally, developers receive feature requests for custom model support, as seen in \textsl{\href{https://github.com/brycedrennan/imaginAIry/issues/255}{imaginAIry \#255\cite{imaginary255}}}.

\textbf{Model API Call} (18 challenges):
Developers often encounter issues when interacting with model providers like OpenAI and Hugging Face, particularly with API calls for remote models. Authorization errors (\textsl{\href{https://github.com/Sanster/lama-cleaner/issues/89}{IOPaint \#89\cite{iopaint89}}}) are the most common. API quota limits also cause disruptions, as seen in \textsl{\href{https://github.com/yoheinakajima/babyagi/issues/32}{babyagi \#32\cite{babyagi32}}}. Also, server connection failures, including bad gateway errors and connection resets, make it difficult to maintain stable communication with model providers.

\textbf{GUI} (16 challenges):
Developers face challenges with GUI malfunctions and usability improvements. Issues like failed model selection, missing input fields, and incorrect defaults cause unexpected behavior (\textsl{\href{https://github.com/NLP-Suite/NLP-Suite/issues/1010}{NLP-Suite \#1010\cite{nlpsuite1010}}}). \rev{Although GUI issues may seem generic, in the context of PTM reuse, they often directly affect model functionality and user control over PTM behavior. For instance, in \textsl{\href{https://github.com/NLP-Suite/NLP-Suite/issues/876}{NLP-Suite \#876\cite{nlpsuite876}}}, a GUI failure prevented users from modifying memory-related configurations, which are essential for executing PTMs. This indicates a strong coupling between GUI elements and PTM-specific resource demands.} Beyond fixing functionality, they also aim to enhance usability specific to PTM workflows, such as intuitive controls and direct model checkpoint selection.

\subsection{Model Performance and Output Quality (PQ)}
Our analysis shows that ensuring high-quality model performance is one of the most frequent challenges (172) that developers face when reusing PTMs. In the following, we elaborate on the sub-themes of the category.

\indent\textbf{Poor Model Output} (106 challenges):
Developers often report missing, inconsistent, or distorted model outputs, especially in image and video tasks like inpainting. Issues include blank images, artifacts, and glitches (\textsl{\href{https://github.com/Sanster/IOPaint/issues/292}{IOPaint \#292\cite{iopaint292}}}). Similar challenges occur in text processing, where outputs are misformatted, exemplified in \textsl{\href{https://github.com/irgolic/AutoPR/issues/65}{AutoPR \#65\cite{autopr65}}}.

\textbf{Model Training and Evaluation} (43 challenges):
This sub-theme covers challenges in model training and fine-tuning, from dataset preparation to evaluation. Developers face issues with invalid or insufficient data, such as incompatible formats and misaligned labels (\textsl{\href{https://github.com/drivendataorg/Zamba/issues/234}{Zamba \#234\cite{zamba234}}}). Training misconfigurations, including unregistered learning rate schedulers and missing validation splits, disrupt workflows. Fine-tuning struggles arise from optimization constraints, lack of built-in support, and inefficient multi-GPU utilization. Evaluation challenges include monitoring gaps, poor visualization tools and inadequate validation metrics, exemplified in \textsl{\href{https://github.com/OpenBB-finance/OpenBB/issues/4136}{OpenBB \#4136\cite{openbb4136}}}. Slow training speeds and caching inefficiencies further limit scalability, highlighting the demand for more efficient pipelines. There are also needs for retraining a model to improve accuracy, as shown in \textsl{\href{https://github.com/drivendataorg/Zamba/issues/54}{Zamba \#54\cite{zamba54}}}.


\textbf{Processing Time} (23 challenges):
Processing speed is another bottleneck in PTM-powered applications, with developers reporting slow or inconsistent inference times, especially in image generation. For example, \textsl{\href{https://github.com/brycedrennan/imaginAIry/issues/113}{ImaginAIry \#113\cite{imaginary113}}} highlights significant performance discrepancies across software versions. Some developers found ineffective processing time in a specific hardware setting, as seen in \textsl{\href{https://github.com/Sanster/IOPaint/issues/270}{IOPaint \#270\cite{iopaint270}}}.

\subsection{Software Environment (SE)}
Developers also frequently encounter difficulties when dealing with software environment around PTMs. This theme includes 118 challenges and is divided into four sub-themes:

\textbf{Dependency Incompatibility} (42 challenges): 
Dependency incompatibility is the most prevalent challenges in this sub-theme, with developers frequently encountering four main types of conflicts: library-to-library, Python version mismatches, library-to-model, and library-to-OS incompatibilities. For instance, strict versioning conflicts are evident in \textsl{\href{https://github.com/Sanster/IOPaint/issues/315}{IOPaint \#315\cite{iopaint315}}}. Similarly, Python version incompatibilities can cause build failures, as reported in \textsl{\href{https://github.com/CogStack/MedCAT/issues/34}{MedCAT \#34\cite{medcat34}}}.

\textbf{Dependency Import and Usage} (28 challenges):
Developers frequently face challenges when importing and using dependencies, particularly PTM-related libraries, like  transformers, as seen in \textsl{\href{https://github.com/brycedrennan/imaginAIry/issues/352}{imaginAIry \#352\cite{imaginary352}}}. They also failed to import other AI-related libraries, such as pytorch and stanza. Beyond import failures, incorrect usage also leads to runtime errors, such as \textsl{\href{https://github.com/JasonKessler/Scattertext/issues/16}{Scattertext \#16\cite{scattertext16}}}.

\textbf{Unsupported or Missing Dependencies} (27 challenges):
Missing or unsupported dependencies hinder model execution, especially AI libraries (e.g., sacremoses, transformers) and system tools, exemplified in \textsl{\href{https://github.com/octimot/StoryToolkitAI/issues/40}{StoryToolkitAI \#40\cite{storytoolkitai40}}}. Hardware acceleration libraries, like CUDA and cuDNN also cause compatibility issues. For example, \textsl{\href{https://github.com/threestudio-project/threestudio/issues/128}{Threestudio \#128\cite{threestudio128}}}, a RuntimeError occurs due to an unsupported OpenGL version.

\textbf{Dependency Installation and Management} (21 challenges):
Managing dependencies through requirements files is essential for maintaining version integrity and preventing inconsistencies. However, developers usually encounter challenges related to package updates, renaming, and installation failures. For instance, package renaming by providers can cause unexpected disruptions, as seen in \textsl{\href{https://github.com/octimot/StoryToolkitAI/issues/47}{StoryToolkitAI \#47\cite{storytoolkitai47}}}. Installation failures are another common issue, particularly when required package versions are unavailable, as reported in \textsl{\href{https://github.com/NLP-Suite/NLP-Suite/issues/47}{NLP-Suite \#47\cite{nlpsuite47}}}.

\subsection{Developer Information and Documentation (DI)}
Developers often struggle with insufficient user guidance, missing or inaccurate documentation, and a lack of essential resources, which hinder usability and accessibility to their model functionality. This theme consists of 115 challenges and is divided into four sub-themes:

\textbf{Needs for User Guidance} (61 challenges):
This sub-theme covers the how-to questions that developers face. A popular question is how to use functionality with user's custom model, as seen in \textsl{\href{https://github.com/neuralchen/SimSwap/issues/245}{SimSwap \#245\cite{simswap245}}}. Similarly, with some software offering the user option for fine-tuning, questions on training and performance optimization arise, such as \textsl{\href{https://github.com/neuralchen/SimSwap/issues/363}{SimSwap \#363\cite{simswap363}}}. Other queries involve model I/O, resource handling, and optimizing models for specific hardware and environments.

\textbf{Generic Documentation} (26 challenges):
Lack of clear and accurate documentation remains a major barrier to developing PTM-based software. Developers often struggle with missing or incomplete guides, as illustrated in \textsl{\href{https://github.com/painebenjamin/app.enfugue.ai/issues/25}{app.enfugue.ai \#25\cite{appenfugue25}}}. Beyond missing resources, inaccurate documentation, including outdated instructions, typos, and inconsistencies, which further complicates the functionality usage for the users.

\textbf{Needs for Other Resources and Information} (18 challenges):
Beyond standard documentation, developers often experience requests for additional resources and technical clarifications to better understand and extend PTM functionality. Common requests include original model training settings, training scripts, backend implementation details, and lists of supported models. For instance, a user requested access to the training file in \textsl{\href{https://github.com/sail-sg/EditAnything/issues/33}{EditAnything \#33\cite{editanything33}}}.

\textbf{Tutorial and Demo} (10 challenges):
Incomplete or inaccurate tutorials and demos also create significant barriers between PTM-related fucntionalities and users. Many struggle with outdated guides, misleading instructions, and missing content for critical tasks such as model usage, training, and inference, as shown in \textsl{\href{https://github.com/CogStack/MedCAT/issues/152}{MedCAT \#152\cite{medcat152}}}.

\subsection{Hardware and Computation Resource (HR)}
Efficient hardware utilization is critical for PTM-applied software, yet developers frequently encounter challenges related to GPU and CPU in terms of memory constraints and ineffective computational resource usage. This theme comprises 115 challenges and is divided into four sub-themes:

\textbf{Memory Usage} (49 challenges):
PTMs are highly memory-intensive, often exceeding hardware limits and causing execution failures. Out-of-memory errors (\textsl{\href{https://github.com/Sanster/IOPaint/issues/3}{IOPaint \#3\cite{iopaint3}}}) and insufficient memory allocation dominate this sub-theme, making model execution infeasible under constrained hardware. Developers also struggle with illegal memory access and the lack of memory optimization, such as \textsl{\href{https://github.com/Rubiksman78/MonikA.I/issues/50}{MonikA.I \#50\cite{monikai50}}}.

\textbf{Hardware Initialization and Execution Management} (30 challenges):
Managing hardware resources poses significant challenges, with developers facing CUDA failures, execution provider issues, and multiprocessing inefficiencies. Missing CUDA devices (\textsl{\href{https://github.com/Sanster/IOPaint/issues/104}{IOPaint \#104\cite{iopaint104}}}) are also common, along with ONNXRuntime CUDA loading failures. Additionally, multiprocessing issues, such as improperly released semaphore objects used for process synchronization, contribute to execution failures.

\textbf{Hardware Incompatibility} (18 challenges):
Hardware mismatches, such as unsupported processing units and suboptimal tensor allocations, create bottlenecks in PTM execution. For example, \textsl{\href{https://github.com/brycedrennan/imaginAIry/issues/300}{imaginAIry \#300\cite{imaginary300}}} reflects a device mismatch issue, where tensors were placed on different devices (MPS and CPU). Some models fail to leverage hardware accelerators effectively due to compatibility issues, such as Pytorch MPS not supporting the specific model (\textsl{\href{https://github.com/Sanster/IOPaint/issues/286}{IOPaint \#286\cite{iopaint286}}}).

\textbf{Computation Utilization} (18 challenges):
Even when hardware is available, balancing CPU-GPU workloads and optimizing resource allocation remain important as the previous one. Developers often report inefficient GPU utilization and poor CPU-GPU assignment strategies, such as \textsl{\href{https://github.com/neuralchen/SimSwap/issues/192}{SimSwap \#192\cite{simswap192}}}.

\subsection{Model Resource and Configuration (RC)}
We also observed challenges related to the management of model resources and configurations. In this theme, 111 challenges are covered and categorized into seven sub-themes:

\textbf{Missing Model Resource} (37 challenges):
Developers often encounter missing model files and model checkpoint files, required for model usage, inference, or resuming training, such as \textsl{\href{https://github.com/CogStack/MedCAT/issues/19}{MedCAT \#19\cite{medcat19}}}. In addition, missing configuration files and text embedding files further complicate model functionality execution, forcing developers to reconfigure or re-download the necessary resources.


\textbf{Inability to Read-Write Model Resource} (21 challenges):
Encoding, serialization, and decompression issues frequently obstruct model reading and writing processes. For example, developers found a Unicode encoding error, as seen in \textsl{\href{https://github.com/neuralchen/SimSwap/issues/288}{SimSwap \#288\cite{simswap288}}}.


\textbf{Model Resource Download} (15 challenges):
Downloading PTM resources is another challenge, with developers reporting HTTP connection failures and authorization restrictions (\textsl{\href{https://github.com/brycedrennan/imaginAIry/issues/12}{imaginAIry \#12\cite{imaginary12}}}). Developers also face inefficient downloading mechanisms, as exemplified in \textsl{\href{https://github.com/nomic-ai/nomic/issues/164}{Nomic \#164\cite{nomic164}}}.

\textbf{Model Resource Storage Management} (13 challenges):
Developers face challenges in locating, organizing, and storing model resource files, including inconsistent directory structures, storage path conflicts, and difficulties with dynamic imports, as shown in \textsl{\href{https://github.com/marieai/marie-ai/issues/20}{Marie-ai \#20\cite{marieai20}}}. Large model files also exceed storage limits, causing checkpoint failures and caching issues, especially in constrained environments like Colab.

\textbf{Model Resource Format and Integrity} (12 challenges):
Developers face compatibility issues when dealing with model formats, structural inconsistencies, and integrity verification failures, as seen in \textsl{\href{https://github.com/neuralchen/SimSwap/issues/173}{SimSwap \#173\cite{simswap173}}} and \textsl{\href{https://github.com/neuralchen/SimSwap/issues/9}{SimSwap \#9\cite{simswap9}}}.

\textbf{Model Configuration Management} (7 challenges):
Handling model configurations remains a challenge, with developers encountering incorrect default configuration parameter settings and lack of flexibility in modifying configurations. Moreover, developers highlight the need for centralized configuration management, as exemplified in \textsl{\href{https://github.com/CogStack/MedCAT/issues/72}{MedCAT \#72\cite{medcat72}}}.

\textbf{Model Resource Accessibility and Security} (6 challenges):
Developers also receive reports from users about denied access to model resources, such as vocabulary files. Additionally, developers encounter security issues. For example, in \textsl{\href{https://github.com/CogStack/MedCAT/issues/65}{MedCAT \#65\cite{medcat65}}}, a developer reported that they suffered spam attacks downloading their model files, causing a service disruption. \rev{A similar case is \textsl{\href{https://github.com/yoheinakajima/babyagi/issues/15}{BabyAGI \#15\cite{babyagi15}}}, mismanagement of API keys, probably creating security risks and unexpected costs.}

\subsection{Model Functionality Enhancement Needs (FN)}
Despite the bugs found during PTM integration, developers usually face enhancement requests. This category has 106 challenges and is divided into five sub-themes:

\textbf{Additional Model Support Needs} (45 challenges):
Developers receive numerous requests for supporting additional models, especially for image-related tasks like generation, embedding, inpainting, and 3D reconstruction. For example, \textsl{\href{https://github.com/brycedrennan/imaginAIry/issues/337}{imaginAIry \#337\cite{imaginary337}}} shows a demand for models with superior performance in specific scenarios. Requests also extend to other domains, such as text (e.g., SPECTER and BERT), audio (e.g., YAMNet), and video (e.g., Whisper).

\textbf{Model Supplementary Functionality Needs} (30 challenges):
Developers also face the demands for developing the supplementary features to the existing core functionality for several purposes. Enhancing user experience is a key priority, with users seeking better interfaces, usability refinements, and more accessible configuration options, such as \textsl{\href{https://github.com/neuralchen/SimSwap/issues/18}{SimSwap \#18
\cite{simswap18}}}. Additional requests focus on optimizing model performance, output quality, and operational efficiency.

\textbf{Model Core Functionality Expansion Needs} (13 challenges):
Aside from supplementary functionalities, they also receive requests to expand the model capabilities to cover additional core features, particularly in image and text-related tasks. These include image-to-text, image-to-3D, outpainting, refinement, and negative prompt-based generation, primarily driven by user demands, such as \textsl{\href{https://github.com/threestudio-project/threestudio/issues/65}{Threestudio \#65\cite{threestudio65}}}. Additional requests include audio and video features, such as non-English speech and multi-language transcription from video.

\textbf{Needs for Incorporation with External Software} (13 challenges):
PTM-applied software often receives requests for third-party integrations to enhance usability and functionality. For example, \textsl{\href{https://github.com/yoheinakajima/babyagi/issues/105}{babyagi \#105\cite{babyagi105}}} suggests integrating SingularGPT for UI automation using natural language. Other requests focus on performance optimization, cost reduction, and cross-platform compatibility.

\textbf{Model Replacement Needs} (5 challenges):
Apart from adding and expanding model capabilities, developers also receive requests to replace existing models, as illustrated in Figure \ref{fig:motivation} and discussed in the motivation section. These replacements are often driven by the need to improve resource efficiency. Other motivations include performance optimization, cost reduction, and version upgrades, as exemplified in \textsl{\href{https://github.com/brycedrennan/imaginAIry/issues/326}{imaginAIry \#326\cite{imaginary326}}}.

\section{Resolution Time (RQ2)}
\label{sect:results_2}

\begin{figure}[!htb]
\centering
\begin{subfigure}[t]{0.46\linewidth} 
    \centering
    \includegraphics[width=\linewidth]{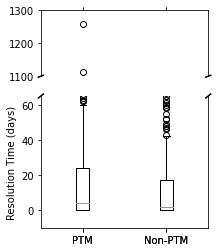}
    \caption{PTM vs. Non-PTM.} 
    \label{fig:ptm_vs_non_ptm}
\end{subfigure}
\hfill
\begin{subfigure}[t]{0.46\linewidth} 
    \centering
    \includegraphics[width=\linewidth]{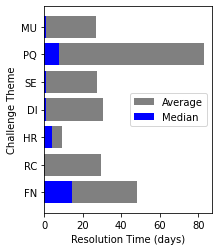}
    \caption{Fix Duration by Theme.} 
    \label{fig2:avg_resolution_time}
\end{subfigure}
\caption{Resolution time analysis of PTM-related issues.}
\label{fig:resolution_time}
\end{figure}

Figure \ref{fig:ptm_vs_non_ptm} presents box plots that compare resolution times for PTM and non-PTM issues. As observed, PTM-related issues take longer to resolve, averaging 40 days compared to 26 days for non-PTM issues. In terms of minimum and maximum resolution times, both groups include issues that were resolved on the same day (0 days). However, PTM-related issues exhibit a much longer maximum resolution time of 1,260 days, compared to 507 days for non-PTM issues. In addition, the Mann-Whitney U Test confirms that PTM-related issues take significantly longer to resolve than non-PTM issues, with a p-value of 0.047. 

Figure \ref{fig2:avg_resolution_time} presents the average and median resolution times across top-level challenge themes. A Kruskal-Wallis H-test reveals a statistically significant difference in resolution time across different challenge categories ($p < 0.001$). This indicates that certain challenge categories take disproportionately longer to resolve than others. \textit{Model Performance and Output Quality (PQ)} has the highest average resolution time (79 days). The longest sub-theme is Model Training and Evaluation, averaging 202 days. Within this sub-theme, Model Training Needs, which includes requirements for model retraining and fine-tuning to improve accuracy, was the most prolonged issue. Note that only the top-level themes are shown in Figure \ref{fig2:avg_resolution_time}.

The findings also show that \textit{Model Functionality Enhancement Needs (FN)} has the second-highest average resolution time, at 48 days. We found that issues under this theme often involve enhancements rather than traditional bug fixes. These issues require developers to assess multiple factors before deciding to implement them, such as implementation complexity, availability of existing models or functionalities, project scope limitations, accuracy concerns, potential risks of misuse, and development costs. Notably, we observed multiple cases where feature requests were rejected outright, not due to technical infeasibility, but rather because they failed to meet the above considerations, as exemplified in \textsl{\href{https://github.com/yoheinakajima/babyagi/issues/326}{babyagi \#105\cite{babyagi105}}}, \textsl{\href{https://github.com/brycedrennan/imaginAIry/issues/116}{imaginAIry \#116\cite{imaginary116}}}, and \textsl{\href{https://github.com/brycedrennan/imaginAIry/issues/132}{imaginAIry \#132\cite{imaginary132}}}.





\section{Discussion}
\label{sect:discussion}

\subsection{Overview of Challenges in PTM Reuse}

From Figure \ref{fig:taxonomy}, out of a total of 915 challenges, we observed that the most frequently discussed PTM-related challenges in downstream software projects are \textit{Model Usage} and \textit{Model Performance and Output Quality}. These findings highlight a critical need for improved handling of model I/O (i.e., the highest number of challenges within \textit{Model Usage}), and addressing issues related to poor-quality model outputs (i.e., the most prevalent among all sub-themes). Conversely, \textit{Model Functionality Enhancement Needs} and \textit{Model Resource and Configuration} cover fewer challenges. Despite this difference, the overall range between the most and least discussed challenges remains relatively balanced, indicating a uniform distribution of difficulties encountered by developers. Furthermore, the high frequency of documentation-related issues highlights the need for clearer and accurate user guidance, which remains insufficient, as reflected in the numerous challenges under \textit{Developer Information and Documentation}. Lastly, even challenges with lower frequencies, such as \textit{Model Resource Accessibility and Security}, should not be overlooked, as neglecting them could potentially lead to significant negative impacts.

\subsection{Taxonomy Comparison}

We compare our taxonomy (RQ1) with existing taxonomies of the four most relevant and recent studies\cite{Humbatova2020, Lai2024, Tan2024, Taraghi2024} to discuss unique and additional challenges found in downstream software projects reusing PTM. For a precise comparison, we create a mapping between the codes in our taxonomy and those in the existing taxonomies at the lowest level \cite{anonymous_2025_15020622}. We analyze their codes based on their explanations, sample issues, and the replication package. Where conceptual overlap existed, we aligned our codes accordingly, with each of their codes mapping to none, one, or multiple codes in our taxonomy, depending on the degree of overlap. The color bars below each box in Figure \ref{fig:taxonomy} show the percentage of overlapping codes where white bars indicate new challenges.  
Overall, we found that most of the codes identified in our study were distinct from existing taxonomies, especially in \textit{Model Functionality Enhancement Needs}, \textit{Model Resource and Configuration}, and \textit{Model Performance and Output Quality}.


Comparing with the challenges in traditional DL/ML-based GitHub projects by Humbatova et al. \cite{Humbatova2020} (red bar) and Lai et al. \cite{Lai2024} (blue bar) in Figure \ref{fig:taxonomy}, their taxonomies extensively cover \textit{Model Usage} issues, particularly in model I/O handling, model initialization, and API interactions as they mainly focus on developing traditional models from scratch. 
Although Lai et al. \cite{Lai2024} further address \textit{Software Environment} challenges, they overlook \textit{Model Resource and Configuration}, which is critical in PTM reuse (e.g., managing model files, checkpoints, and embeddings), leading to storage challenges. The challenges of \textit{Developer Information and Documentation}, \textit{Model Performance and Output Quality} issues (e.g., fine-tuning difficulties) and prompting challenges in \textit{Model Usage} were not found in traditional AI-based projects. This highlights the increased complexity of PTM-applied software, where both development and usage require extensive technical support and resources.


For the taxonomies of PTM reuse mined from  Stack Overflow discussions by Tan et al. \cite{Tan2024} (yellow bar) and model repositories on Hugging Face by Taraghi et al. \cite{Taraghi2024} (green bar) in Figure \ref{fig:taxonomy}, the complexities of PTM reuse in real-world downstream projects were not captured. 
A remarkable omission is \textit{Model Functionality Enhancement Needs}. Unlike other model users, downstream software projects demand additional model support, model replacement, core and supplementary functionality expansion, and third-party software integration. As discussed in our resolution time analysis, these enhancement requests require developers to evaluate feasibility, complexity, and software constraints before implementation, highlighting the increased effort required when reusing PTMs. 
Many challenges were also not captured in prior taxonomies~\cite{Tan2024, Taraghi2024}.
For example, GUI-related challenges as PTM-powered applications frequently incorporate GUIs to enhance usability.
Although both taxonomies include \textit{Developer Information and Documentation} challenges, the unclear documentation when using PTMs in specific applications was not captured.
Other aspects such as poor model output, processing time bottlenecks, and computational utilization issues also were not captured in the prior taxonomies, highlighting the unique challenges of PTM reuse which are far beyond model selection and adaptation.



\subsection{PTM-related Resolution Time}

For the discussion on resolution time, the findings from RQ2 underscore that PTM issues take longer to be resolved than non-PTM issues. A possible reason behind this is that many PTM issues go beyond simple bug identification and fixing. Instead, they involve model performance optimization, which was found to have the highest average fix time. This process is typically more complex and iterative. Debugging these issues often requires retraining models or modifying data pipelines, tasks that demand considerable time. \textit{Model Functionality Enhancement Needs} could be another contributing factor, as we observed that issues under this theme often depend on multiple aspects, requiring developers to invest additional effort to resolve them. Interestingly, we observed that while \textit{Model Usage} is the most prevalent challenge developers face, issues under this category tend to be resolved more quickly than most other themes, with the exception of \textit{Hardware and Computation Resource}, which has the shortest fix time.


\section{Implications}
\label{sect:implication}

 \subsection{Implications for Practitioners}

Understanding the challenges can help downstream software developers anticipate pitfalls, refine development practices, and implement proactive strategies. We recommend the followings: {\large \textcircled{\normalsize 1}} \textbf{Enhancing Model Functionality Performance \& Customizability}—Based on RQ1, integrating monitoring tools could help streamline PTM training and improve model performance and output quality. Supporting lightweight model-based functionality (e.g., \cite{Kriman2020, victor2019, Howard2017, Gandhi2023}) and multi-processing options would enable users to run models with limited resources. Additionally, developers should implement support for custom or user-specified models (e.g., from Hugging Face or local storage) via command-line prompting. {\large \textcircled{\normalsize 2}} \textbf{Improved Documentation \& Usability Support}—As indicated by our RQ1 findings, user guidance issues are one of the dominant challenges. Developers should provide comprehensive and versioned documentation, especially clear usage instructions (e.g., prompting, CLI, GUI, hardware requirements) and ensure alignment with evolving dependencies. Additionally, we suggest that developers employ automated tools \cite{Moreno2017, Gao2023} to simplify technical documentation, thereby enhancing user accessibility. {\large \textcircled{\normalsize 3}} \textbf{Streamlining Resource Management \& Configuration}—To address the challenges in \textit{Model Resource and Configuration}, particularly the sub-theme with the longest fix duration—Model Resource Management, standardized configuration formats (e.g., YAML \cite{filip2024}) may help centralize settings in a single file, making management easier. Implementing model resource caching can minimize redundant downloads, improving efficiency for end users. {\large \textcircled{\normalsize 4}} \textbf{Strategic Resource Allocation for PTM Issue Resolution}—From RQ2, PTM-related issues take longer to resolve, particularly in model training issues and requests for model functionality enhancements. Developers can benefit from allocating extra effort, prioritizing tasks by expected resolution time \cite{Lai2024}, and documenting common failures. Moreover, clear acceptance criteria \cite{wallace2019} may also help manage enhancement requests.

\subsection{Implications for Researchers}
Our systematic analysis of PTM reuse challenges can help researchers identify current gaps, inspire new research directions, and develop novel solutions. Based on our findings, we propose the following research directions: {\large \textcircled{\normalsize 1}} \textbf{Automated PTM Issue Resolution \& Effort Estimation}—The results of RQ2 indicate that PTM-related issues vary in complexity and resolution time. We suggest researchers to explore automated tools for prioritizing \cite{Bugayenko2023}, estimating effort\cite{Choetkiertikul2019}, and predicting resolution duration \cite{banyongrakkul2023} specifically for PTM issues, aiding developers in better resource allocation. {\large \textcircled{\normalsize 2}} \textbf{Enhancing Feature Request Evaluation \& PTM Evolution}—Based on the findings of RQ2, many requests from \textit{Model Functionality Enhancement Needs} are declined due to implementation complexity, project scope limitations, or security concerns. This raises key research questions: How can software engineering practices better support PTM evolution? What factors influence the adoption and replacement of PTMs in OSS projects, and how do developers decide which models to support or replace? \cite{Jiang2024} What trade-offs exist between expanding PTM functionality and maintaining software stability? {\large \textcircled{\normalsize 3}} \textbf{Lightweight PTM Development \& Hardware Efficiency}—Given the resource-intensive nature of PTMs, highlighted by the prominence of memory usage issues in RQ1, we suggest that researchers explore techniques to optimize model efficiency, including lightweight architectures \cite{Long2024}, quantization and hardware-aware model design \cite{Dilhara2021} to improve accessibility across various computing environments. {\large \textcircled{\normalsize 4}} \textbf{Addressing PTM Compatibility \& Environment Challenges}—Incompatibilities between PTM-related dependencies hinder smooth integration. Future research could focus on automated tools for conflict detection and resolution, alternative library suggestions, and improved compatibility mechanisms, as emphasized in existing work \cite{Dilhara2021, Tan2024}.





\section{Threats to Validity}
 \label{sect:threats} 
 
\subsection{Internal Validity}
Filtering software projects and PTM-related issue reports have not been previously studied in the literature. To reduce bias, we define clear inclusion/exclusion criteria and employ two annotators for manual classification \rev{with inter-annotator agreement measurement. In the first annotation phase, which involved filtering software projects, the agreement score was 0.65. Although this may appear moderate, it is generally considered a substantial level of agreement based on the guidelines by Landis et al. \cite{Landis1977}, and aligns with accepted standards in software engineering research \cite{Elberzhager2012}. In the second phase, which focused on identifying PTM-related issues, the initial agreement was slightly lower at 0.60, but improved significantly to 0.82 in the second round after a reconciliation discussion.} While manual challenge categorization for RQ1 involves some subjectivity \cite{Ratner2002}, we follow a rigorous approach \cite{Cruzes2011, Braun2006}, including a pilot study for annotator familiarization and iterative refinement through weekly discussions. Additionally, resolution time in RQ2 may be misleading as an issue report can contain multiple challenges, probably skewing the distribution. However, such cases are limited, and we address this by using non-parametric tests to assess median differences, reducing sensitivity to outliers. 

\subsection{External Validity} 
Our base dataset, PeaTMOSS \cite{Jiang2024}, is externally sourced which might introduce unknown biases. However, we mitigate potential bias through multiple rigorous filtering steps, including manual annotation with agreement checks. Additionally, in comparative analysis, some prior studies exhibit inconsistencies between the taxonomies described in their papers and the resources provided in their replication packages, while others lack detailed coverage of software environment issues. To address this, we thoroughly examine all available issue report examples. Furthermore, to enhance transparency, we include the comparative evidence in our replication package \cite{anonymous_2025_15020622}.

\subsection{Construct Validity} 
\rev{Our study aims to identify challenges specifically related to the reuse of PTMs in downstream software projects. However, there is a risk that a few identified sub-themes may appear generic to general software systems (e.g., GUI). To mitigate this threat, we focus our analysis on how these challenges manifest themselves uniquely in the context of PTM reuse. In RQ2, r}esolution time may not always reflect actual fix effort, as compounding factors such as issue type, developer workload, and limited interaction with the reporter may influence it \cite{Banyongrakkul2024}. To mitigate this threat, we controlled the distribution for two significant factors—project and issue type—by \rev{employing stratified random sampling when selecting non-PTM issues to match the PTM group. Both groups were drawn from the same annotated dataset, and the matching process ensured a balanced distribution across projects and issue types. To validate the reliability of the issue type classification by seBERT used for stratification, we also performed a broader manual validation of a representative sample to indicate that the use of seBERT introduces minimal risk of systematic bias. While this reduces sampling bias and improves comparability, we acknowledge that residual confounding factors and sensitivity to sample variation may still influence results.} We also accounted for extreme outliers by using non-parametric tests, ensuring our analysis is less sensitive to anomalies. 

\section{Conclusion}
\label{sect:conclusion} 
This paper presents a comprehensive and systematic analysis of the challenges derived from PTM-related issue reports found in downstream OSS GitHub projects. We manually conducted rigorous data filtering to create a curated dataset of 840 PTM-related issue reports across 31 OSS projects reusing PTMs. Our data analysis reveals seven main themes of challenges: Model Usage, Model Performance and Quality, Software Environment, Developer Information and Documentation, Hardware and Computation Resource, Model Resource and Configuration, and Model Functionality Enhancement Needs. We also found that PTM-related issues take significantly longer to resolve than non-PTM issues.

Based on the taxonomy comparisons, we found that the majority of our identified challenges remain unaddressed in existing work, emphasizing the novelty of our findings. Existing studies primarily focus on Model Usage and Software Environment issues, while our taxonomy introduces new areas in Model Functionality Enhancement Needs, Model Resource and Configuration, and Model Performance and Output Quality. We also provided practical recommendations for downstream developers and researchers to improve the practices in this field, highlighting the need for more robust tooling, better documentation, systematic strategies in software development with PTMs, and future research directions.

\bibliographystyle{IEEEtran}
\bibliography{references}
\end{document}